\documentclass[rnote]{aa} 
\usepackage{graphicx}
\usepackage{txfonts}
\usepackage{natbib}
\begin{document}

   \title{Remarks on the properties of elliptical galaxies in modified Newtonian dynamics}

   \author{T. Richtler
          \inst{1}
          \and
          B. Famaey
          \inst{2,3}
          \and 
          G. Gentile
          \inst{4}
          \and 
          Y. Schuberth
          \inst{2}
          }

   \institute{Universidad de Concepci\'on, Departamento de Astronomia, Concepci\'on, Chile
         \and
             Argelander Institut f\"ur Astronomie, Universit\"at Bonn, Bonn, Germany
         \and
             Observatoire Astronomique de Strasbourg, CNRS, Strasbourg, France
          \and 
              Sterrenkundig Observatorium, Universiteit Gent, Gent, Belgium}

   \date{Received ...; accepted ...}
 
  \abstract
   {Two incorrect arguments against MOND in elliptical galaxies could be that the equivalent circular velocity curves tend to become flat at much larger accelerations than in spiral galaxies, and that the Newtonian dark matter halos are more concentrated than in spirals. }
   {Here, we compare published scaling relations for the dark halos of elliptical galaxies to the scaling relations expected for MONDian phantom halos.}
   {We represent the baryonic content of galaxies by spherical profiles, and their corresponding MONDian phantom halos by logarithmic halos. We then derive
   the surface densities, central densities, and phase space densities and compare them with published scaling relations.}
   {We conclude that it is possible to get flat circular velocity curves at high acceleration in MOND, and that this happens for baryonic distributions described by Jaffe profiles in the region where the circular velocity curve is flat. Moreover, the scaling relations of dark halos of ellipticals are remarkably similar to the scaling relations of phantom halos of MOND.}
  {}


   \maketitle
%

\section{Introduction}

While data on large scale structures point towards a Universe dominated by dark matter and dark energy, e.g. \citet{komatsu11}, the nature of these is still a deep mystery \citep[e.g.,][]{frieman08,wiltshire08,bertone10,kroupa10}. In this context, it is good to keep in mind that this conclusion essentially relies on the 
assumption that gravity is correctly described by Einstein's General Relativity in the extreme weak-field limit, a regime where the need for dark matter itself prevents the theory from being tested. Until this double dark mystery is solved, it is thus worth investigating alternative paradigms and their implications.

For instance, Modified Newtonian dynamics \citep[MOND]{milgrom83} naturally explains various {\it spiral} galaxy scaling relations \citep{tully77, mcgaugh00, mcgaugh04}. The existence of a very tight baryonic Tully-Fisher relation for disk galaxies \citep{mcgaugh05, trachternach09} is for instance one of the remarkable predictions of MOND. The corresponding relation for early-type galaxies is much more
difficult to investigate because they are pressure-supported systems, and the equivalent circular velocity curves determined from the velocity dispersion profiles suffer from the well-known degeneracy with anisotropy. However, some studies circumvented this problem: for instance, \citet{krona00} used data on 21 elliptical galaxies to construct non-parametric models from which circular velocity curves, radial profiles of mass-to-light ratio, and anisotropy profiles as well as high-order moments could be computed. This led \citet[hereafter G01]{gerhard01} to publish benchmark scaling relations for ellipticals. It was e.g.  shown for the first time that circular velocity curves tend to become flat at much larger accelerations than in spiral galaxies. This would seem to contradict the MOND prescription, for which flat circular velocities typically occur well below the acceleration threshold $a_0 \sim 10^{-8}$~cm~s$^{-2}$, but not at accelerations of the order of a few times $a_0$ as in ellipticals. Also \citet[hereafter T09]{thomas09} published scaling relations for dark matter halos of 18 Coma galaxies, using similar prescriptions
as G01. We remark that G01 employed spherical models while the models of T09 are axisymmetric.

Not many studies have considered the predictions of MOND in elliptical galaxies. \citet{milgrom84} showed that pressure-supported isothermal systems have finite mass in MOND with the density at large radii falling approximately as $r^{-4}$. It was also shown that there exists a mass-velocity dispersion
relation of the form $(M/10^{11}M_\odot) \approx (\sigma_r/100\,\,
{\rm kms}^{-1})^4$ which is similar to the observed Faber-Jackson relation \citep{sanders2000, sanders2010}, and that, in order to match the fundamental plane, MOND models must deviate from being strictly isothermal and isotropic: a radial orbit anisotropy in the outer regions is needed \citep{sanders2000,cardone11}. \citet{tiret07} and \citet{angus08} also analyzed the distribution of velocity dispersion of PNe on scales of 20 kpc, and of satellites on very large scales of the order of 400~kpc around red isolated ellipticals, showing that MOND allowed to fit both scales successfully. 

Hereafter, we make general remarks on the properties of  spherical galaxies within MOND, and their scaling relations. We first point out a remarkable property of elliptical galaxies exhibiting a flattening of their circular velocity curve at small radii: such a flattening in the intermediate gravity regime is actually generated by a baryonic density distribution following a Jaffe profile in these parts of the galaxies. We then further show that the observational scaling relations for the dark halos of the elliptical galaxy sample by G01 are strikingly similar to the theoretical ``phantom'' halos of MOND (i.e. the halo that would produce in Newtonian gravity the same additional gravity as MOND), with one remarkable exception: MOND predicts that the product of the central density with the core radius should be constant, as recently observed for spiral galaxies \citep{donato09, gentile09}.

\section{Flat circular velocity curves and the Jaffe profile}

Although it has been argued that some ellipticals do not need any dark matter or enhancement of gravity \citep{aaron03}, there are many counter-examples \citep{magorrian, richtler04, schuberth06, kumar07}. Such a recent example is the elliptical galaxy NGC 2974 where the presence of an HI disk allowed a more or less direct measurement of circular velocities \citep{weijmans08}. There is also evidence that elliptical galaxies exhibit flat circular velocity curves , but that, contrary to spiral galaxies, this happens in the {\it inner regions} where $g>a_0$ (e.g., G01, Weijmans et al. 2008).  Such a flattening of circular velocities is {\it a priori} not expected in the strong to intermediate gravity regime in MOND, and poses the question of how to analytically interpret it. 

In the intermediate gravity regime, the transition from Newtonian to MONDian dynamics is described by the $\mu$-function of MOND. Many concordant studies have recently shown that, in spiral galaxies, the ``simple'' transition of \citet{fb05} is a good representation of the data \citep*[for an extensive discussion]{gentile11}. In a spherical system, with this simple transition, the enclosed (baryonic) mass $M_M(r)$ needed to produce the same gravitational potential in MOND as the (baryonic+dark) mass $M_N(r)$ in Newtonian gravity is:
\begin{equation}
M_M (r) = M_N(r) - \left( \frac{1}{M_N(r)} + \frac{G}{r^2a_0} \right)^{-1}.
\end{equation}
In a region where the circular velocity is constant $v_c=V$ (even if $g>a_0$), one can write $M_N(r)= V^2r/G$, and thus after some algebra
\begin{equation}
M_M(r) = \frac{V^4}{a_0 G} \cdot \frac{r}{r+V^2/a_0}.
\end{equation}
Remarkably, this enclosed mass profile corresponds precisely to a Jaffe profile \citep{jaffe83} with scale-radius $r_j=V^2/a_0$ (meaning that the acceleration is $a_0$ at $r_j$),
and with total mass 
$M_{\rm tot}=V^4/(a_0 G)$. 
Indeed, as the enclosed mass $M_M(r) = M_M(r_0) + 4 \pi \int_{r_0}^r{\rho(R) R^2 dR}$, this enclosed mass profile corresponds locally to the density profile:
\begin{equation}
\rho(r) = \frac{M_{\rm tot}}{4 \pi} \cdot \frac{r_j}{r^2~ (r+r_j)^2},
\end{equation}
with the characteristic surface density (see also Milgrom 1984) $M_{\rm tot}/r_j^2=a_0/G$. This profile is of course not valid for the very inner parts of an elliptical galaxy, where $V$ is not constant. Let us also note that (i) it was already known that a Jaffe profile produces a flat circular velocity curve at $r \ll r_j$ in Newtonian gravity, which MOND generalizes to radii $r \sim r_j$; (ii) $M_{\rm tot}$ does not necessarily have to be the real total mass of the galaxy, as the Jaffe profile fit to the density distribution could have a cut-off in the outer parts. In that case, the  constant circular velocity $V$ would actually fall slightly above the prediction from the baryonic Tully-Fisher relation of spiral galaxies.
Interestingly, this is precisely what is observed for the G01 sample  
of ellipticals.

The fact that elliptical galaxies can exhibit (equivalent) circular velocity curves that are flat in the intermediate gravity regime is thus analytically understood in MOND by the fact that the outer regions of ellipticals can be approximated  by a Jaffe profile with a large scale-radius, i.e. in regions well within the intermediate gravity regime rather than in the deep-MOND regime. These flat circular velocity curves would have been impossible with exponential density profiles (as encountered in
spiral galaxies), meaning that the fact that circular velocity curves become flat quicker in ellipticals does not come as a surprise in the context of MOND.

This finding looks like an interesting possibility to devise new tests of MOND based on photometry. However,
in reality it might be difficult: not many spherical galaxies with a precisely measured density profile are
dynamically investigated out to large radii, and have enough tracers to measure the higher order moments and constrain the anisotropy. Moreover, light might not trace the baryonic mass precisely. As an example, the circular velocity in NGC 2974, which can be traced by an HI disk, becomes constant at around 5 kpc and
has the value 300~${\rm kms}^{-1}$, which would correspond to a Jaffe scale radius of 23 kpc.
Unfortunately, NGC 2974 is neither spherical nor does its photometry reach
large radii so that it does not serve well as a test object. In any case, Weijmans et al.~(2008, their Fig.~20) showed that the reverse procedure (going from the density to the circular velocity curve) leads to a very good fit.

\section{Dark matter scaling relations for phantom halos of ellipticals}

\begin{table*}[t]
\caption{The table shows  for baryonic Hernquist profiles with mass and effective radius described by the first two columns, the corresponding parameters of the MONDian ``phantom'' halo
represented by a logarithmic potential fitted from the center to two effective radii of the baryonic profile. The columns are the baryonic mass, the luminous effective radius, the core radius $r_0$ and the asymptotic velocity $v_0$ of the
log-halo, its surface density, central density and phase space density, the latter as defined by G01.The last column column gives the acceleration in units of 
$a_0$ at a radius of 2 $R_{eff}$ for each Hernquist model. The predictions for MONDian halos in the previous columns are valid  
only for galaxies embedded in an external field smaller than this  
value.}
\begin{center}
\begin{tabular}{cccccccc}
baryonic mass $[M_\odot]$ &$ R_{eff}$ [kpc] & $r_0$ [kpc] & $v_0$[${\rm kms}^{-1}$]  & $S$$[M_\odot/pc^2]$ & $\rho_0$$ [M_\odot/pc^3]$ & $f_{ps}$& acc.[$a_0$]\\
\hline
   $10^{12}$     &  14.1 &  8.83 &   244 &  374 &  0.04    &   8.82$\times$ $10^{-9}$  & 1.53\\
          $8\times10^{11}$     &  11.8 &  7.60 &   228 &  379 &  0.05    &   $1.19\times 10^{-8}$ & 1.55                                \\
    $5\times 10^{11}$     &    8.06 &  5.20   & 193    &  393 & 0.076    & $ 3.02\times  10^{-8}$   & 1.91             \\ 
    $2\times10^{11}$         &     3.84 & 2.87 &  146    &  411   & 0.143    &  $1.3\times 10^{-7}$   & 2.96     \\
    $10^{11}$         &     1.47 & 1.24 &  99    &  438   & 0.35    &  $1.02\times 10^{-6}$  & 8.44\\
    $5 \times10^{10}$      & 1.25     &   1.04      &  90          &  431    &  0.41  &$1.6\times 10^{-6}$ & 6.08          \\
\hline    
\end{tabular}
\end{center}
\label{surfdens}
\end{table*}

We now apply the reverse procedure, and check whether the phantom halos predicted by the simple transition of MOND (Famaey \& Binney 2005) comply with the observational scaling relations of dark halos of ellipticals. As stated above, Jaffe profiles are not good descriptions of the very inner parts of ellipticals. We hereafter rather choose Hernquist profiles \citep{hernquist90} to represent the baryonic content of ellipticals: these are realistic enough and allow for an exhaustive exploration of their properties without varying too many free parameters. Such a Hernquist-model is described by  its total mass $M$ and scale-radius $r_{H}$.
The profile of the Newtonian circular velocity curve then reads 
\begin{equation}
v_N(r) = \sqrt{\frac{G M r}{(r+r_H)^2}}
\end{equation}
 where
the scale-radius $r_H$ of the Hernquist-model is related to the effective (half-light) radius by $ R_{eff} = 1.815 \, r_H$. 

Adopting the simple transition formula between the Newtonian and the MONDian regime, one
finds for the MOND circular velocity
\begin{equation}
 v_{M} = \sqrt{v_N^2(r)/2 + \sqrt{v_N^4(r)/4 + v_N^2(r)  a_0 r}},
\end{equation}
and the MONDian phantom halo has the circular velocity
\begin{equation}
v_{\rm phantom} =  \sqrt{\sqrt{v_N^4(r)/4 + v_N^2(r)  a_0 r}-v_N^2(r)/2}
\end{equation}

To enable the comparison with the scaling relations of G01, where the dark matter halos are adopted as logarithmic halos, we  fit $v_{\rm phantom}(r)$ to  the circular velocity $v_{\rm log}(r)$ of a logarithmic halo with asymptotic circular velocity $v_0$ and core radius $r_0$:
\begin{equation}
 v_{\rm log}(r) = v_0 r/\sqrt{r_0^2+r^2}.
 \end{equation}
The fits are performed within the inner two effective radii \footnote{Let us note that these fits are not particularly good: the circular velocity curve $v_c(r) = v_0 r/(r_0+r)$ would have provided better
fits, but the core radius of the corresponding halo would then be systematically smaller with respect to G01}. The fitted central dark matter density is then given by 
\begin{equation}
\rho_0 = 3 (v_0/r_0)^2/(4 \pi G).
\end{equation}
The characteristic central phase space density is defined (see G01) as 
\begin{equation}
f_{ps} = 3^{3/2} \rho_0/v_0^3.
\end{equation}
The characteristic surface density within $r_0$ is then also defined as (see also Donato et al. 2009):
\begin{equation}
S = \rho_0 \cdot r_0. 
\end{equation}
Table \ref{surfdens} lists these fitted parameters for six baryonic Hernquist masses over a large mass range. The combinations of the masses and effective radii  in Table \ref{surfdens}  follow  equation (5) of G01 (in accordance with the fundamental plane), where we transformed their luminosities into masses by using   $M/L_B$=8  for all galaxy baryonic masses.  

\begin{figure}[]
\begin{center}
\includegraphics[width=0.4\textwidth]{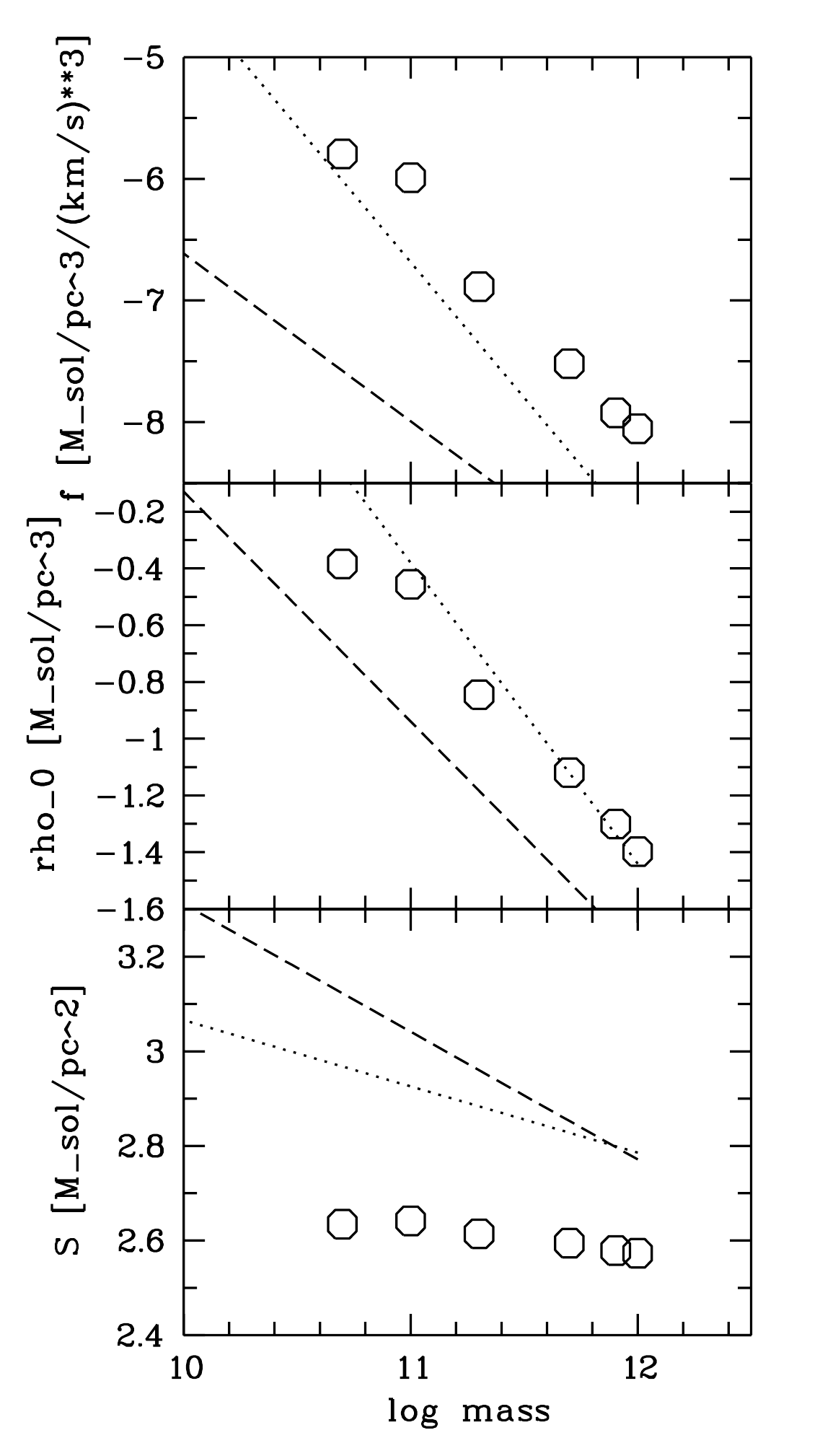}
\caption{The figure shows the surface density, central density, and the central phase space density logarithmically (see the text for
more explanations) of the phantom dark halos (circles) for different baryonic Hernquist masses from Table \ref{surfdens} together with the relations  given by G01 (dotted lines) and T09 (dashed lines). Note that these parameters observationally exhibit a very large scatter around the mean relations. Note also that the relations of T09 are not given explicitly in their paper but have been constructed  from their Table 3 omitting
galaxies with young stellar cores. }
\label{fig:gerhard}
\end{center}
\end{figure}

Fig.\ref{fig:gerhard} shows the values of the fitted dark halo parameters derived from applying MOND to the baryonic Hernquist-profiles, together with the observational scaling relations given by G01 (dotted lines) and T09 (dashed lines). The upper panel shows the characteristic phase space density, the middle panel the central volume density, and the lower panel the characteristic surface density.
Let us note that the plotted relations are indicative only, since the data (Fig.18 of G01 and Figs. 1 and 4 of T09) show a very large scatter even when logarithmically displayed. However, within this observational uncertainty, it is remarkable that some features are perfectly reproduced, particularly the slopes of the  phase space density and of the central volume density as a function of baryonic mass (given the observational scatter, the almost perfect reproduction of the central volume density of G01 might of course be {\it partly} coincidental).

As first emphasized by G01, the phase-space density values are at a given mass higher than in spirals, which means that under the $\Lambda$ Cold Dark Matter paradigm, dark halos of ellipticals cannot be the result of collisionless mergers of present-day spirals, but must have been assembled at a very early time, when the cosmological density was higher. In MOND this is of course not necessarily the case, as the phase-space argument does not apply to phantom halos.

One also notes a remarkable exception to the scaling relations: the fitted characteristic dark matter surface density $S$ is fully independent from the Hernquist parameters, and it is systematically lower than in G01 and T09. We emphasize that this constancy is not related to the special relation of mass and effective radius.  Varying $R_{eff}$ by a factor of two at a given mass does not change  the constant  surface density significantly. This prediction of MOND thus brings the value closer to the (also constant) value of $S$ observed in spiral galaxies, ${\rm log}S=2.1$ \citep{donato09}. Let us note that MOND also predicts the observed constant value of $S$ in spirals, which is somewhat lower because (i) spirals are a bit deeper into the MOND regime  (\citealt{milgrom09}) and (ii) their flattened baryonic profiles lead to a somewhat higher Newtonian gravity at a given mass, and in turn a somewhat lower MOND contribution to the phantom halo. 

\begin{figure}[]
\begin{center}
\includegraphics[width=0.4\textwidth]{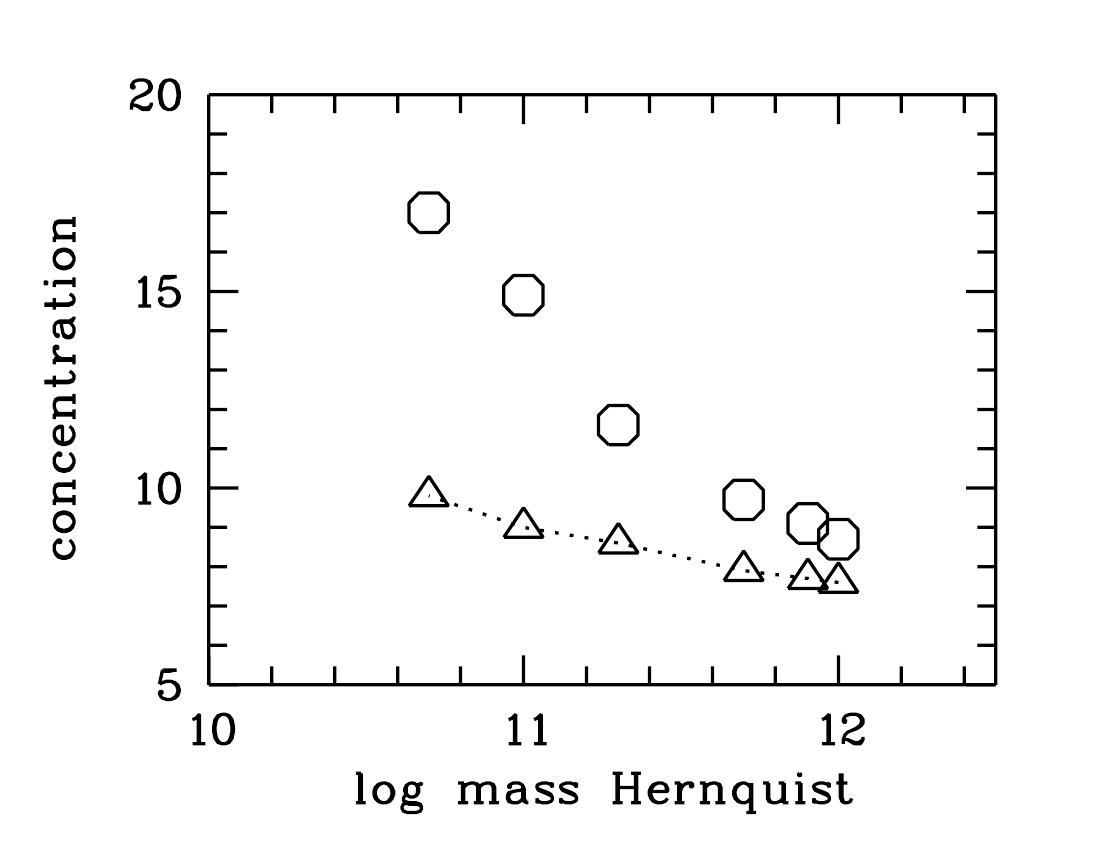}
\caption{The figure shows for our 6 Hernquist masses the concentration parameters of the associated NFW-halos, if the MONDian phantom halos
are fitted by NFW profiles (open circles). The triangles are the concentration parameters expected from the relation quoted by \citet{maccio08}}
\label{fig:M_C}
\end{center}
\end{figure}

On the first glance one might interpret this  constancy and the other scaling relations  as a clear signature of MOND in 
ellipticals: however, CDM may also predict that the surface density within the scale radius  
of NFW halos weakly depends on dark matter total mass \citep{boyarsky10}.  
 For spiral galaxies, this is of little interest as it is known that cuspy profiles often do not fit rotation curves \citep{fb05, deblok10, gentile05}, the mystery then being how to erase the cusp by the feedback from the baryons while keeping the product $ \rho_0 r_0$ constant. In elliptical galaxies, the situation is less clear as NFW profiles often do fit the data equally well as cored profiles \citep{schuberth10}. We thus fitted NFW profiles to the same MONDian phantom halos
 and found a perfect agreement. The question remains whether these NFW-halos are ''cosmological'' or in other words, fulfill the relation between virial mass and concentration predicted by cosmological simulations. Fig.\ref{fig:M_C} displays for our Hernquist masses the resulting concentrations of the NFW-halos (open circles) corresponding to the MONDian phantom halos, while the triangles show the concentration values expected from the equation (9) of \citet{maccio08}, using  200 times the critical density as the mean density within the virial radius (standard cosmology: h=0.7, $\Omega_m = 0.3$, $\Omega_\Lambda=0.7$).
 One concludes that for high masses the MONDian phantom halos are not distinguishable from cosmological NFW halos, given also that the simulations predict considerable scatter. For smaller masses the difference between MONDian phantom halos and NFW
cosmological halos is larger. 

\section{External field effect}
Due to the non-linearity of MOND and its associated breaking of the  
Strong Equivalence Principle, a MONDian stellar system embedded in an  
external gravitational field (EF) stronger than its own internal field  
behaves in a quasi-Newtonian way, with an effectively higher  
gravitational constant 
\citep{milgrom83,famaey07}. Most of the sample galaxies  are located in clusters or groups where the EF might
have an influence. \citet{wu10} for instance showed how the EF can lead to the  
lopsidedness of an originally axisymmetric non-isolated galaxy. 

\begin{figure}[]
\begin{center}
\includegraphics[width=0.4\textwidth]{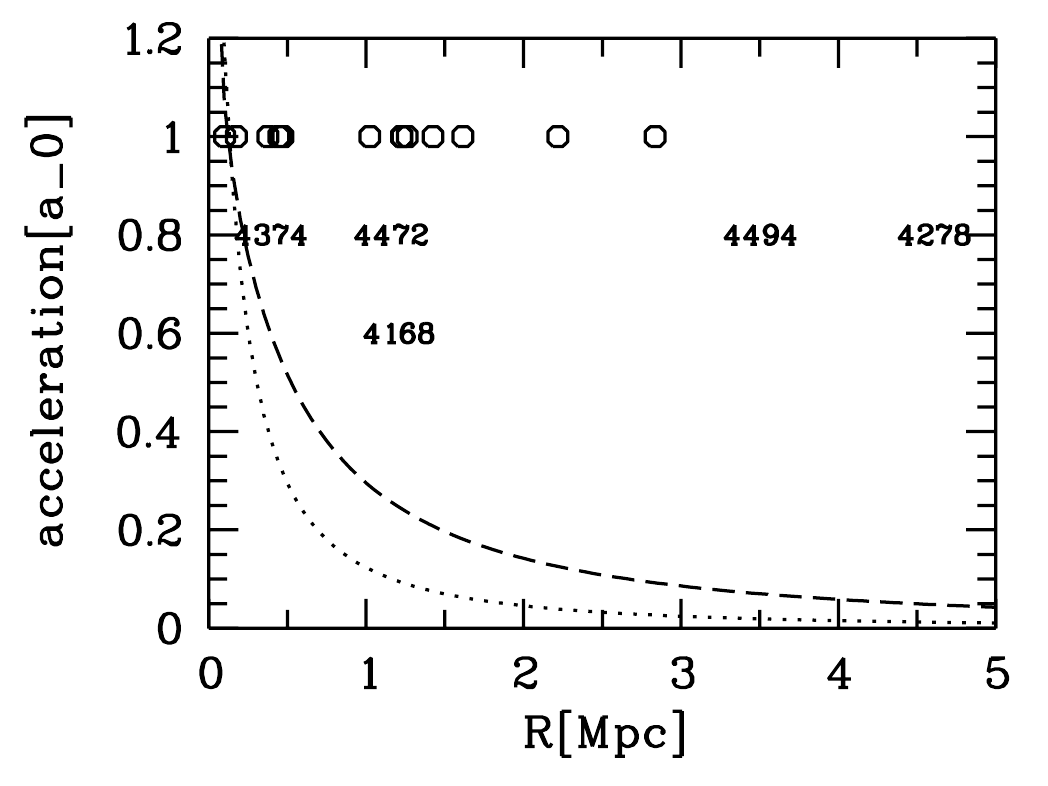} 
\caption{This plot estimates the external field acting on Virgo (dashed) and Coma (dotted) galaxies.  Abscissa is the projected distance in Mpc from M87 and
NGC 4874, respectively. Ordinate is the
acceleration in units of $a_0$.  Adopted distances for Virgo and Coma are 15 Mpc and 100 Mpc, respectively. The values for Virgo are generated by an extrapolation  of the mass model  for M87 of \citet{mclaughlin99}.  Some G01 galaxies are indicated by their NGC numbers. 
The values for Coma are generated by using the NFW dark halo from \citet{lokas03}. Small open circles are the 18 T09 galaxies whose
projected distances are taken from \citet{godwin83}.
The comparison with Table \ref{surfdens} shows that the EF is small.}
\label{EF}
\end{center}
\end{figure}

While it is beyond the scope of this research note to evaluate in detail the EF in the present sample, a very rough estimation is presented in Fig. \ref{EF}, which plots for the  
Virgo and the Coma cluster the accelerations based on the extrapolations of the mass models cited in the figure caption (these extrapolations are only meant to give an order of magnitude  
estimate, but should not be taken as rigorous models). This can be compared with the internal accelerations at 2 $R_{eff}$ for
the Hernquist models in Tab.\ref{surfdens}. Indicated are the projected distances of  galaxies in the Virgo and
Coma region. 
The positions of the Virgo galaxies correspond to the middle points of their NGC numbers, while the Coma
galaxies are plotted as small open circles. 
One concludes that the EF should have no influence in the two  samples at the galactocentric distances which we consider.

\section{Conclusion}

Here we showed that (i) in MOND, galaxies exhibit a flattening of their circular velocity curve at high gravities ($g>a_0$) if they are described by a Jaffe profile with characteristic surface density $a_0/G$ in the region where the circular velocity is constant (since this is not possible for exponential profiles, it is remarkable that such flattenings of circular velocity curves at high accelerations are only observed in elliptical galaxies); (ii) the phantom halos of ellipticals predicted by MOND (i.e., the dark halos that would produce in Newtonian gravity the same additional gravity as MOND) can be fitted by logarithmic halos which perfectly reproduce the observed scaling relations of ellipticals for phase-space densities and central volume densities $\rho_0$; (iii) these halos have a constant characteristic surface density $\rho_0r_0$; (iv) contrary to spirals (for which there are more data in the very central parts), the phantom halos of ellipticals can as well be fitted by cuspy NFW halos, the concentration of which is in accordance with the theoretical predictions of $\Lambda$CDM for the highest masses, but in slight disagreement for baryonic masses smaller than $10^{11} M_\odot$: a modern, large, sample of elliptical galaxies, which are dynamically well investigated out to large radii and cover a large range of masses, will thus be required to get discriminating power. But in any case, and whatever the true physical reason for it, it is remarkable that a recipe (MOND) known to fit rotation curves of spiral galaxies with remarkable accuracy also apparently predicts the observed distribution of ``dark matter'' in
elliptical galaxies.

\begin{acknowledgements}
We thank an anonymous referee for a thoughtful report. 
TR acknowledges financial support from the Chilean Center for Astrophysics,
FONDAP Nr. 15010003,  from FONDECYT project Nr. 1100620, and
from the BASAL Centro de Astrofisica y Tecnologias
Afines (CATA) PFB-06/2007.  BF acknowledges the support of the Humboldt foundation. GG is a postdoctoral researcher of the FWO-Vlaanderen (Belgium).
\end{acknowledgements}

\bibliographystyle{aa}
\bibliography{researchnote.bib}
\end{document}